\documentclass[a4paper,english,aps,prb,amsmath,amsfonts,superscriptaddress, twocolumn,showkeys,dvipsnames]{revtex4-1}
\usepackage{lmodern}
\usepackage{lmodern}
\usepackage[latin9]{inputenc}
\usepackage{babel}
\usepackage{array}
\usepackage{amstext}
\usepackage{graphicx}
\usepackage{booktabs}
\usepackage{multirow}
\usepackage[unicode=true,
 bookmarks=false,
 breaklinks=false,pdfborder={0 0 1},backref=false,colorlinks=false]
 {hyperref}

\makeatletter

\DeclareFontEncoding{LGR}{}{}
\DeclareRobustCommand{\greektext}{%
  \fontencoding{LGR}\selectfont\def\encodingdefault{LGR}}
\DeclareRobustCommand{\textgreek}[1]{\leavevmode{\greektext #1}}
\ProvideTextCommand{\~}{LGR}[1]{\char126#1}

\newcommand{\lyxmathsym}[1]{\ifmmode\begingroup\def\b@ld{bold}
  \text{\ifx\math@version\b@ld\bfseries\fi#1}\endgroup\else#1\fi}

\DeclareTextSymbolDefault{\textquotedbl}{T1}
\providecommand{\tabularnewline}{\\}

\pdfoutput=1
\usepackage{multirow}

\IfFileExists{microtype.sty}{\usepackage{microtype}}{\relax}
\def\0{\phantom{0}}

\usepackage{xcolor}

\usepackage{leftindex}

\makeatother

\begin{document}
\title{Lattice vibrational modes in changchengite from Raman spectroscopy
and first principles electronic structure}
\author{B. Chatterjee}
\affiliation{Department of Theoretical Physics, Jozef Stefan Institute, Jamova
39, 1000 Ljubljana, Slovenia}
\author{D. Vengust}
\affiliation{Department of Complex Matter, Jozef Stefan Institute, Jamova 39, 1000
Ljubljana, Slovenia}
\author{A. Mrzel}
\affiliation{Department of Complex Matter, Jozef Stefan Institute, Jamova 39, 1000
Ljubljana, Slovenia}
\author{P.Sutar}
\affiliation{Department of Complex Matter, Jozef Stefan Institute, Jamova 39, 1000
Ljubljana, Slovenia}
\author{E. Goreshnik}
\affiliation{Dept. of Inorganic Chemistry and Technology, Jozef Stefan Institute,
Jamova 39, 1000 Ljubljana, Slovenia}
\author{J. Mravlje}
\affiliation{Department of Theoretical Physics, Jozef Stefan Institute, Jamova
39, 1000 Ljubljana, Slovenia}
\author{T. Mertelj}
\email{tomaz.mertelj@ijs.si}

\affiliation{Department of Complex Matter, Jozef Stefan Institute, Jamova 39, 1000
Ljubljana, Slovenia}
\affiliation{Center of Excellence on Nanoscience and Nanotechnology Nanocenter
(CENN Nanocenter), Jamova 39, 1000 Ljubljana, Slovenia}
\date{\today}
\begin{abstract}
We measured room-temperature phonon Raman spectra of changchengite
(IrBiS) and compared the experimental phonon wavenumbers to the theoretical
ones obtained by means of the \emph{ab initio} density-functional-theory
calculations in the presence and absence of the spin-orbit coupling
effects. Combining two different excitation photon energies all the
symmetry predicted Raman modes are experimentally observed. The electronic
properties of IrBiS are found to be similar to the recently studied
isostructural compound IrBiSe showing a large Dresselhaus spin-orbit
valence band splitting. A good agreement between the experimental
and theoretically predicted Raman phonon wavenumbers is found only
when the lattice parameter is constrained to the experimental value.
The inclusion of the spin orbit coupling does not significantly affect
the phonon wavenumbers.
\end{abstract}
\maketitle

\section{Introduction}

\label{sec:intro}

Changchengite with chemical formula IrBiS is a naturally occurring\cite{zuxiang1997changchengite}
pyrite (FeS$_{2}$) related compound, but was first reported as an
artificially synthesized compound\citep{hulliger1963new} and found
to be semiconducting as expected from the chemical similarity to pyrite.
Recently, its sister compound IrBiSe has been highlighted for its
large spin-orbit splitting resulting in fully spin-polarized valence-band
pockets with a peculiar 3D chiral spin texture. It was suggested\citep{liu2020giant}
that the presence of such pockets could offer a possibility for generation
of electrically controlled spin polarized currents for spintronic
applications.

\emph{Ab initio} calculations\citep{zhang19} for IrBiSe also suggest
a strong bulk photo voltaic effect and a large shift-current. Combination
of symmetry analysis and \emph{ab initio} calculations\cite{liu2021charge}
for IrBiSe further predict a realization of Weyl phonons characterized
by Chern number $\pm$ 4.

One expects similar behavior also in IrBiS, however, apart from the
initial discovery structural and transport data\citep{hulliger1963new}
and recent ARPES results\citep{liu2020giant} in IrBiSe virtually
no other experimental data exist for IrBiS and IrBiSe. Contrary to
IrBiSe the electronic structure of IrBiS has not been explored.

In the present study we measured phonon Raman spectra in IrBiS single
crystals and compared them to theoretical predictions by \emph{ab
initio }density-functional methods taking into account the strong
spin-orbit coupling (SOC). At first, neglecting the SOC, we find a
decent agreement between the experimental and theoretically calculated
Raman phonon wavenumbers. Inclusion of the SOC does not significantly
improve the agreement unlike observed in the previous studies in the
heavy metals Bi and Pb, where the inclusion of SOC significantly improved
the agreement.\citep{diaz2007phonon,verstraete2008density} On the
other hand, constraining the lattice parameter to the experimental
value further improved the agreement between theory and experiment.

The theoretical electronic band structure of IrBiS is found to be
similar to IrBiSe. The reasonably good agreement between the \emph{ab
initio} density functional calculations and experimental electronic
band structure in IrBiSe by ARPES\citep{liu2020giant} and the experimental
lattice dynamics in IrBiS in the present case demonstrate that band-structure
calculations work well in this class of compounds and could be used
to further explore suitability of related compounds for spintronic
applications.

For comparison we also calculated the Brillouin zone center phonon
modes of IrBiSe, which agree to those reported in Ref. {[}\onlinecite{liu2021charge}{]}
and are, as expected, softer compared to IrBiS.

The paper is organized as follows: We describe the experimental procedure
in Section II and the computational methods in Section III. In Section
IV we present and discuss the experimental and theoretical results.
Finally, in Section V we summarize and conclude. In Appendix we present
our theoretically calculated phonon mode wavenumbers in IrBiSe and
compare it to the phonon mode wavenumbers in IrBiS.

\begin{figure}
\includegraphics[width=1\columnwidth]{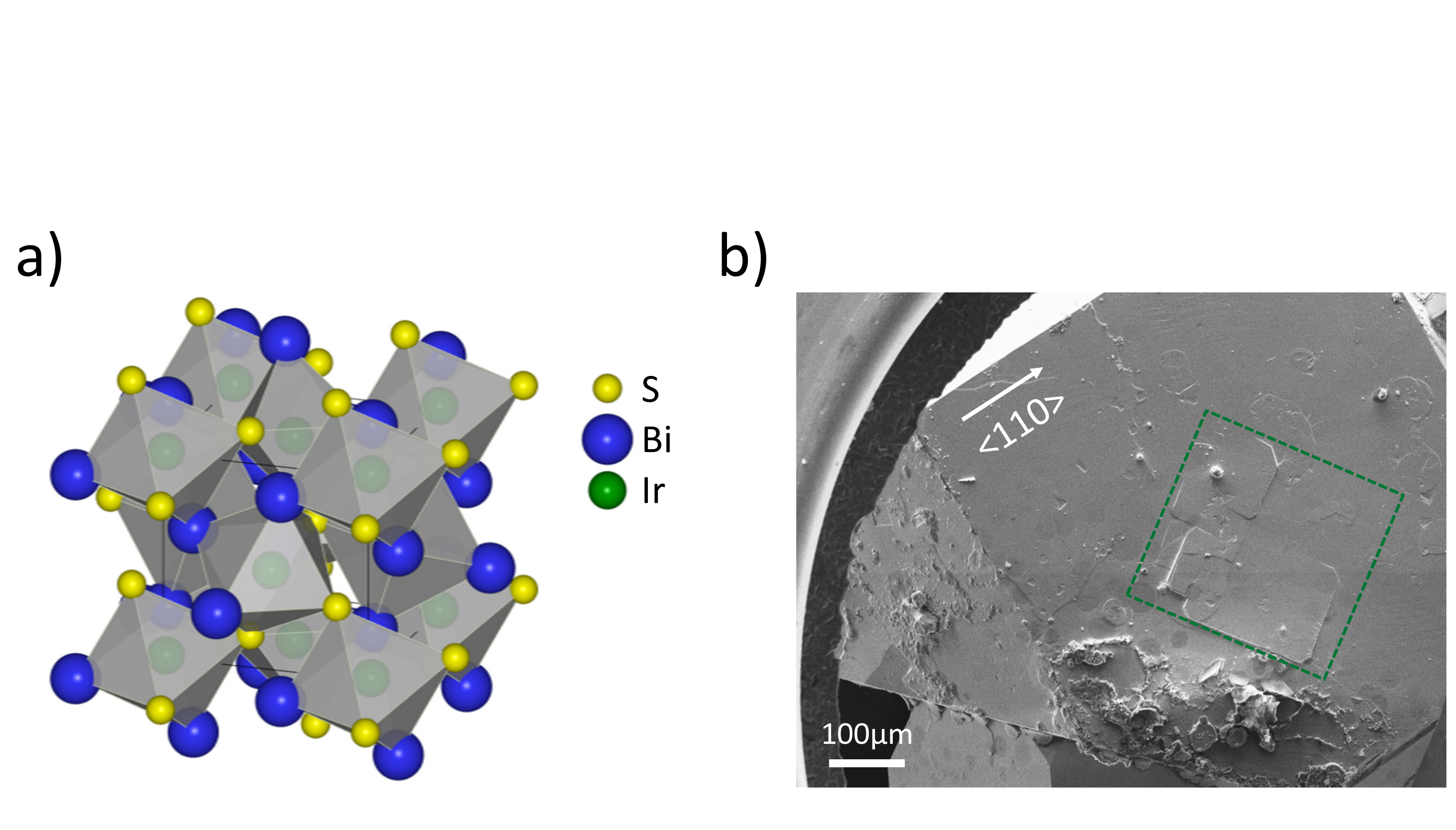}\caption{\label{fig:struct}(a) Schematics\citep{momma2011vesta} of IrBiS
cubic P2$_{1}$3 crystal structure. Each Ir ion is coordinated by
a combination of three Bi and three S ions forming distorted octahedron.
Each of Bi and S ions is shared by three such octahedra. (b) Scanning
electron image of the largest crystal. The axes orientation was determined
from the morphology and confirmed by the Raman selection rules. Some
impurities and a $\sim45{}^{\circ}$rotated overgrowth (green rectangle)
are observed in the bottom part.}
\end{figure}

\section{Experimental}

\subsection{Crystal growth and characterization}

IrBiS crystallizes in pyrite derived non-centrosymmetric simple cubic
crystal structure (space-group P2$_{1}$3 (No. 198), point group T
(23)) with 4 formula units in the unit cell ($a_{\mathrm{exp}}=6.143$
Å).\citep{hulliger1963new} The structure can be viewed as a network
of distorted IrBi$_{3}$Se$_{3}$ octahedra with each Bi or Se ion
shared to three such octahedra {[}see Fig. \ref{fig:struct} (a){]}.

IrBiS crystals were grown unintentionally during growth of thiospinel
CuIr$_{2}$S$_{4}$ single crystals from Bi flux\citep{naseska2020orbitally}
following the procedure developed in Ref.{[}\onlinecite{matsumotoNagata2000}{]}.
In the present case 0.196 g of CuIr$_{2}$S$_{4}$ powder was mixed
with 6.5 g of Bi and vacuum sealed in a double compartment\citep{matsumotoNagata2000}
quartz tube. The tube was put into a furnace programmed to ramp to
1050$^{\circ}$C in 24h with a 72h dwell followed by slow cooling
of 1.5 $^{\circ}$C/h to 500 $^{\circ}$C. At this temperature most
of the liquid phase is usually separated from the solid phases by
appropriately tilting the tube followed by 12h cooling to a room temperature.
The resulting crystals are cleaned of the remaining solidified flux
by washing in hot concentrated HNO$_{3}$.

During the crystal growth producing IrBiS crystals the furnace was
found to be at room temperature with the program stopped 12 days after
the program has started, presumably due to a transient power outage
sometime between day 2 and 12 or programming error. The heat treatment
procedure was restarted from the beginning using the same sample/quartz
tube containing the partially processed solidified melt. After finishing
the procedure by washing the product in hot concentrated HNO$_{3}$
a clump of a few IrBiS single crystals was recovered instead of the
usual CuIr$_{2}$S$_{4}$ single crystals. Due to the described anomaly
during the initial heat treatment run the particular growth conditions
for IrBiS single crystals are unfortunately not known in detail and
a controlled synthesis/growth of IrBiS single crystals has not been
yet achieved.

Single-crystal X-ray diffraction data from the obtained crystals were
collected on a Gemini A diffractometer equipped with an Atlas CCD
detector, using graphite monochromated Mo-K$_{\alpha}$ radiation.
The collected diffraction data were processed with the CrysAlis PRO
program. For unit cell checking 6 runs of 30 frames were collected.
The unit cell belonging to the cubic system with the lattice constant
$a=6.14$ {Å} was determined using 98 observed reflections. The
found unit cell fully corresponds to that found in ICSD database for
BiIrS (rec. No. 616740)\citep{hulliger1963new}.

A scanning electron microscope image of the largest obtained IrBiS
crystal is shown in Fig. \ref{fig:struct} (b). The crystal shows
$\left\langle 100\right\rangle $ and $\left\langle 111\right\rangle $
facets and composition Ir:Bi:S close to 1:1:1 (see Tab. SI (Supporting
Information)). Impurities present as chunks on otherwise smooth surfaces
are different phases containing also Cu and were not studied in detail.

\subsection{Raman scattering}

Room temperature Raman scattering experiments were conducted in a
micro-Raman setup (back-scattering configuration) using 632.8 nm and
488 nm excitation using 10x optical objective and $\sim1$ mW laser
power. Possible sample damage due to the laser excitation was checked
by measuring a few spectra at lower as well as higher (using 20x and
60x objectives) excitation power densities and no differences or deterioration
were observed with increasing excitation power density.

The spectra were collected from as grown $\left\langle 100\right\rangle $
facets where the excitation and detection polarizations were oriented
according to the facet edge morphology resulting in systematic mode
extinctions as expected from the group theoretical analysis (see also
Subsection \ref{subsec:Raman-scattering}).

\section{Computational methods}

\label{sec:method}


We performed electronic structure calculations using the density functional
theory (DFT) by means of the pseudo-potential plane-wave Quantum ESPRESSO
software package that we also used to perform structural relaxations
and to compute the phonon modes at the $\Gamma$-point using the density
functional perturbation theory (DFPT).\citep{giannozzi2017quantum}
The basis set cut-off for the plane-wave, and charge density expansion
was chosen to be 55 Ry and 440 Ry respectively. The Brillouin zone
was sampled on a 6 by 6 by 6 uniform grid according to the Monkhorst-Pack
scheme. The threshold parameter\citep{giannozzi2017quantum} \textquotedbl tr2\_ph\textquotedbl{}
controlling the convergence of the self-consistent calculation of
the phonon modes using DFPT was set to $10^{-15}$. Calculations performed
with a larger threshold parameter resulted in nonphysical/negative
phonon wavenumbers due to uncompleted convergence.

The exchange-correlation functional was treated within the generalized
gradient approximation (GGA) using Perdew-Burke-Ernzerhof (PBE) functional.
The SOC was included using a fully relativistic ultra soft pseudo
potential calculated by Dal Corso \citep{dal2014pseudopotentials}.
For the calculations without the SOC we used the ultra-soft non-norm
conserving pseudo-potential, with scalar relativistic correction only,
from the 'SSSP-1.0-PBE-precision' library\citep{sssplib}.

We performed a structural relaxation such that the forces on each
atom in the unit cell were less than $10^{-4}$ Ry/au, and the total
stress was below 
$2\times10^{-7}$ Ry/Bohr$^{3}$. 
The lattice parameter, $a=6.213$~Å\ (volume of the unit cell $V=239.80$~${\lyxmathsym{\AA}}^{3}$),
of the relaxed structure as well as the experimental lattice parameter,
$a_{\mathrm{exp}}=6.143$ Å, were used for the electronic structure
and phonon mode calculations.

We also performed DFT calculations using all-electron, linear augmented
plane-wave basis method as implemented in WIEN2k package \citep{wien2k}
with the following parameters: the radii of the muffin-tin spheres
were $R_{{\rm MT}}({\rm Bi})=2.5\,a_{{\rm B}}$ for bismuth atoms
and $R_{{\rm MT}}({\rm Ir})=2.5\,a_{{\rm B}}$ for iridium atoms,
and $R_{{\rm MT}}({\rm S})=2.0\,a_{{\rm B}}$ for sulfur atoms. The
Brillouin zone was sampled with 10000 k points (451 k points in the
irreducible wedge) in the self-consistent cycle. The basis-set cutoff
$K_{{\rm max}}$ was defined with $R_{{\rm MT}}({\rm Bi})\times K_{{\rm max}}=7.0$.
Scalar relativistic effects as well as the SOC on a second variational
level were included. The band structures obtained from the Quantum
ESPRESSO and WIEN2k packages are consistent. The band structure plots
shown in the present paper were calculated using the WIEN2k package.

\section{Results and discussions}

\subsection{Raman scattering\label{subsec:Raman-scattering}}

Optical modes for IrBiS with non-centrosymmetric space group P2$_{1}$3
are classified according to the irreducible representations of point
group T(23) as\citep{KroumovaAroyo2003} $\text{\textgreek{G}}\mathrm{_{optic}}=3\mathrm{A}+3\leftindex[]^{1}{\mathrm{E}}+3\leftindex[]^{2}{\mathrm{E}}+8\mathrm{T}$.
The label A refers to the non degenerate totally symmetric representation.
$\leftindex[]^{1}{\mathrm{E}}$ and $\leftindex[]^{2}{\mathrm{E}}$
are one dimensional complex conjugate representations that correspond
to two degenerate (due to the time reversal symmetry) modes. T refers
to a triply degenerate representation. All optical modes are Raman
active\citep{KroumovaAroyo2003} while only the T modes are infrared
active. Since the symmetries of all three atomic sites are identical
the modes of each symmetry come in multiples of 3, where one of the
9T triply degenerate groups corresponds to the acoustic branches.

To assign the observed modes symmetry we search for systematic extinctions
as a function of the relative orientations between the excitation
light, the detection analyzer polarizations and the crystal orientation
(in the back-scattering geometry). The A modes Raman tensor\citep{KroumovaAroyo2003}
is diagonal and isotropic with $\chi_{xx}=\chi_{yy}=\chi_{zz}$. The
A modes are therefore extinct in the crossed polarizations configuration
(CPC) irrespective of the crystal orientation. The two E modes Raman
tensors are also diagonal, but anisotropic, $\chi_{xx}\neq\chi_{yy}\neq\chi_{zz}$.\footnote{$\chi_{zz}=0$ for the $\leftindex[]^{2}{\mathrm{E}}$ modes.}
When the excitation polarization is parallel to the $\left\langle 100\right\rangle $
directions the scattered light polarization is parallel to the excitation
one, so the modes are extinct in the CPC, while for a general crystal
orientation the observation is allowed in both, the CPC and the parallel
polarization configuration (PPC). The T modes Raman tensors are purely
off-diagonal and symmetric\footnote{Assuming the absence of strong resonant effects.}
with $\chi_{yz}=\chi_{xz}=\chi_{xy}$. As a result, the T modes are
extinct in the PPC when the the excitation polarization is parallel
to the $\left\langle 100\right\rangle $ directions and in the CPC
when the excitation polarization is parallel to the $\left\langle 110\right\rangle $
directions.

Taking the $z\left(x,x\right)$-$z$ scattering geometry\footnote{In Porto notation.}
the T modes must be systematically extinct, while the A and E modes
must be systematically extinct in the $z\left(x,y\right)$-$z$ scattering
geometry. To separate out the E modes the $z\left(x',y'\right)$-$z$,
with $x'=x+y$ and $y'=x-y$, scattering geometry is instrumental
since both, the A and T modes must be extinct.

\begin{figure}
\includegraphics[width=0.7\columnwidth]{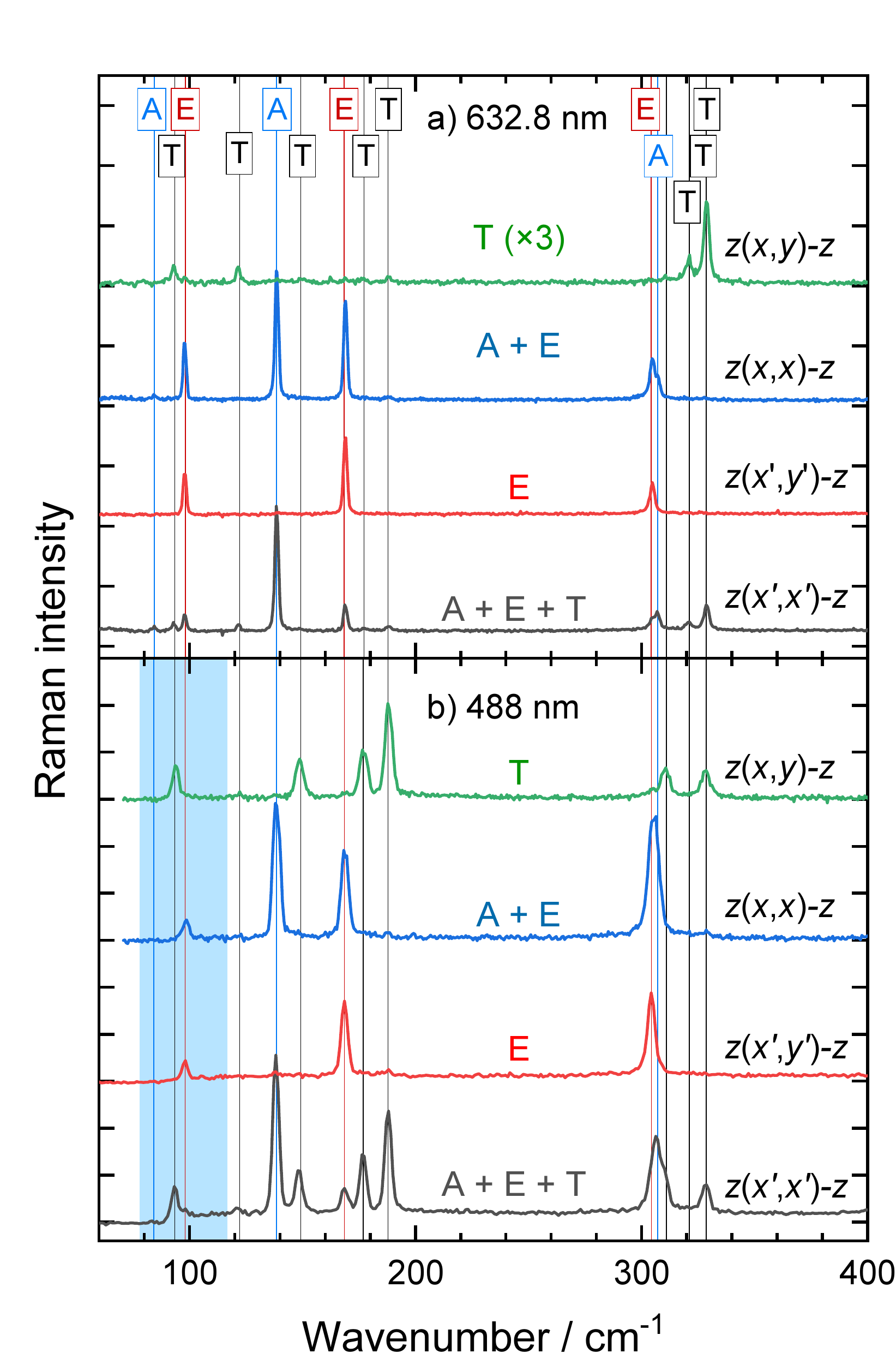} \caption{\label{fig:ramaexpt}Measured Raman spectra in different scattering
geometries a) and b) with 632.8 nm and 488 nm excitation, respectively.
Porto notation is used with $x'=x+y$ and $y'=x-y$. The lines in
b) appear broader due to a worse instrumental resolution at 488 nm.
The blue shaded region in b) corresponds to the region where the sensitivity
strongly drops with decreasing wavenumber due to the 488-nm edge filter.}
\end{figure}

In Fig. \ref{fig:ramaexpt} we show experimental polarized Raman spectra.
The relative intensities of most of the modes are quite different
comparing the two excitation wavelengths indicating the presence resonant
Raman effects. This is expected since both excitation photon energies
are well above the gap.

Combining the spectra at both laser excitation wavelengths we observe
all the symmetry-predicted optical modes where the T modes at 148.5
cm$^{-1}$and 320.9 cm$^{-1}$ are observable only at a single laser
wavelength. The lowest wavenumber (84.4 cm$^{-1}$) A mode is also
missing at 488-nm excitation presumably due to its wavenumber being
at the very edge of the 488-nm laser-rejection filter window. Taking
the morphologically determined crystal orientation all the expected
mode extinctions are observed. The wavenumbers and symmetries of all
modes are compiled in Tab. \ref{tab:phonon_modes}.

We observe a separate group of peaks ($\mathrm{1A}+\mathrm{1\leftindex[]^{1,2}{\mathrm{E}}}+3\mathrm{T}$)
above 300 cm$^{-1}$, which can be associated with the modes dominated
by the lightest S ions displacements. On the other hand, the other
modes observed below 200 cm$^{-1}$ appear as a single group. Since
the masses of Ir and Bi are very similar these modes are expected
to have mixed displacements. Nevertheless, the progression of the
mode symmetries still reflects the mechanical-representation structure,
where each atomic site generates ($\mathrm{1A}+\mathrm{1\leftindex[]^{1,2}{\mathrm{E}}}+3\mathrm{T}$)
representations.\footnote{In the lowest wavenumber group the 3T acoustic modes are not visible
in Raman spectra.}

\begin{table*}[p]
\caption{\label{tab:phonon_modes}Comparison between experimental and calculated
phonon wavenumbers of at the $\Gamma$-point in IrBiS without and
with SOC using the ultra-soft non-norm conserving GGA pseudo potential.}

\begin{ruledtabular}
\begin{tabular}{cccccccccc}
\multicolumn{2}{c}{Experimental} & \multicolumn{3}{c}{Calculations (relaxed, $a=6.213$ {Å})} & \multicolumn{3}{c}{Calculations (constr., $a_{\mathrm{exp}}=6.143$ {Å})} & Symmetry & Activity\tabularnewline
\hline 
632.8 nm & 488 nm & without SOC & with SOC & $\Delta$\footnote{Using the theoretical results with the relaxed lattice constant and
SOC.} & without SOC & with SOC & $\Delta$\footnote{Using the theoretical results with the experimental lattice constant
and SOC.} &  & \tabularnewline
\hline 
\multicolumn{4}{c}{Wavenumber (cm$^{-1}$)} & \% & \multicolumn{2}{c}{Wavenumber (cm$^{-1}$)} & \% & - & -\tabularnewline
\hline 
- & - & -0.20 & 0.0 & - & 0.0 & 0.0 & - & T & acoustic\tabularnewline
84.4 & - & 79.04 & 78.4 & -7.1 & 81.4 & 81.3 & -3.7 & A & R\tabularnewline
93.0 & 93.7 & 83.8 & 84.8 & -9.2 & 86.2 & 87.1 & -6.7 & T & I+R\tabularnewline
97.9 & 98.2 & 91.2 & 92.6 & -5.6 & 95.1 & 96.4 & -1.7 & E & R\tabularnewline
121.6 & 122.0 & 114.1 & 113.7 & -6.7 & 118.4 & 118.3 & -2.9 & T & I+R\tabularnewline
138.6 & 138.4 & 127.6 & 127.5 & -7.9 & 134.4 & 134.9 & -2.6 & A & R\tabularnewline
-- & 148.5 & 133.1 & 135.1 & -9.0 & 142.2 & 145.0 & -2.4 & T & I+R\tabularnewline
169.0 & 168.7 & 154.1 & 155.5 & -7.9 & 163.0 & 164.4 & -2.6 & E & R\tabularnewline
177.1 & 176.9 & 156.2 & 157.4 & -11.1 & 164.4 & 165.9 & -6.3 & T & I+R\tabularnewline
188.1 & 188.0 & 165.9 & 167.2 & -11.1 & 176.4 & 178.1 & -5.3 & T & I+R\tabularnewline
304.8 & 305.4 & 275.9 & 285.6 & -6.4 & 300.1 & 310.0 & 1.6 & E & R\tabularnewline
307.5 & 307.0 & 294.4 & 288.8 & -6.0 & 306.9 & 301.8 & -1.8 & A & R\tabularnewline
311.2 & 310.7 & 282.6 & 290.7 & -6.5 & 306.3 & 313.8 & 0.9 & T & I+R\tabularnewline
320.9 & - & 300.8 & 302.9 & -5.6 & 321.1 & 319.6 & -0.4 & T & I+R\tabularnewline
328.8 & 328.3 & 314.2 & 312.8 & -4.8 & 331.7 & 335.1 & 2.0 & T & I+R\tabularnewline
\end{tabular}
\end{ruledtabular}

\end{table*}

\subsection{Electronic band structure\label{sec:theory_results}}

\label{subsec:bandstructure} 
\begin{figure}[h]
\includegraphics[width=1\columnwidth]{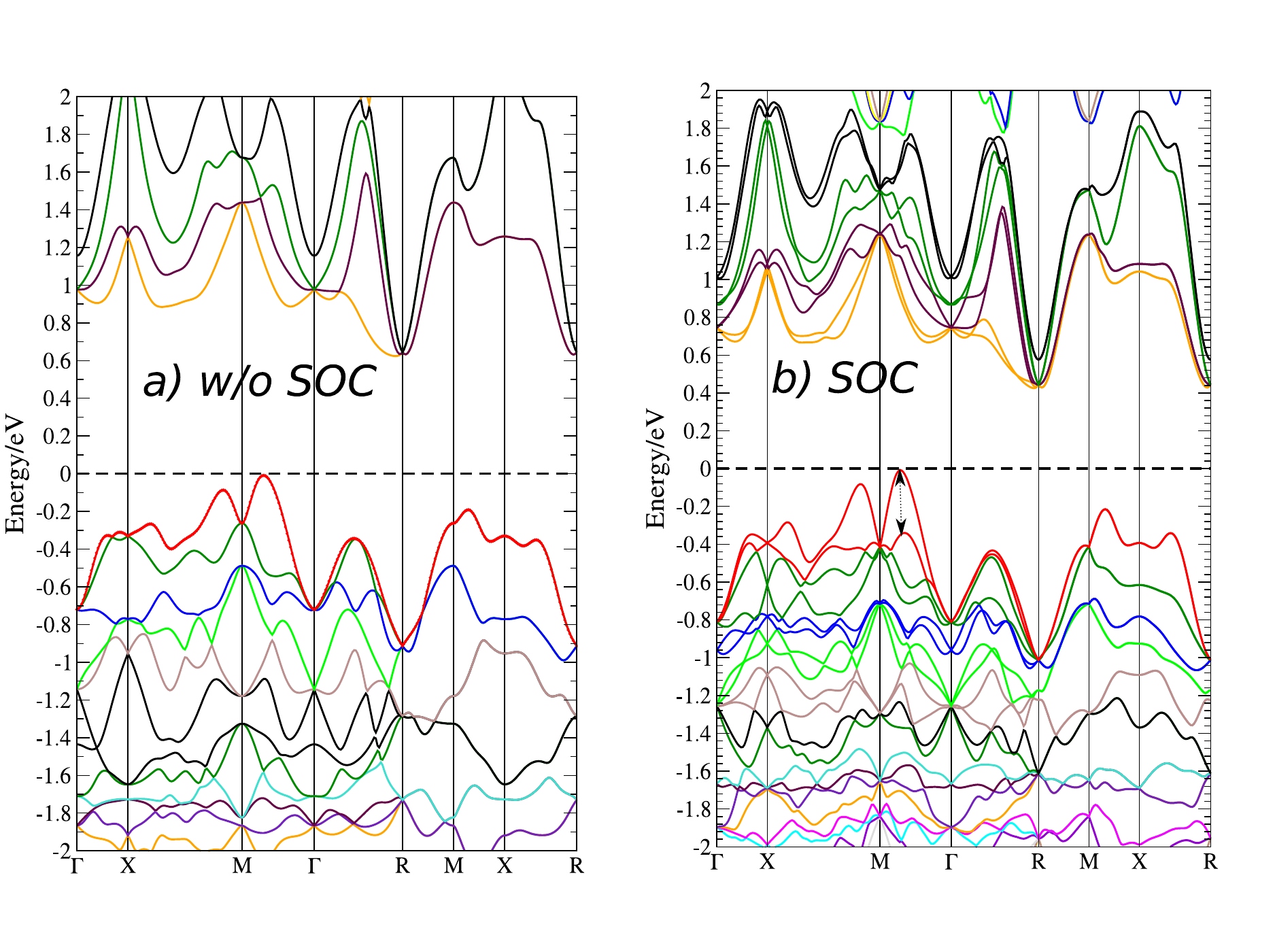} \caption{\label{fig:bands}The electronic band-structure of IrBiS without (a)
and with (b) SOC from the Wien2k calculations using GGA. The dotted
line with a double-headed arrow indicates the large valence band ($\sim0.34$-eV)
Dresselhaus splitting due to the SOC. The dashed line at 0 eV corresponds
to the ($T=0$ K) Fermi level.}
\end{figure}

In Fig. \ref{fig:bands} we present
a comparison of the electronic band structure without (a) and with
(b) SOC using the relaxed lattice parameters. A clean band-gap appears
across the Fermi level indicating that IrBiS is a band-insulator/semiconductor.
We find an overall agreement of the band structure\citep{liu2020giant}
and the SOC induced effects on the bands with the related isostructural
compound IrBiSe.

\begin{figure}[h]
\includegraphics[width=1\columnwidth]{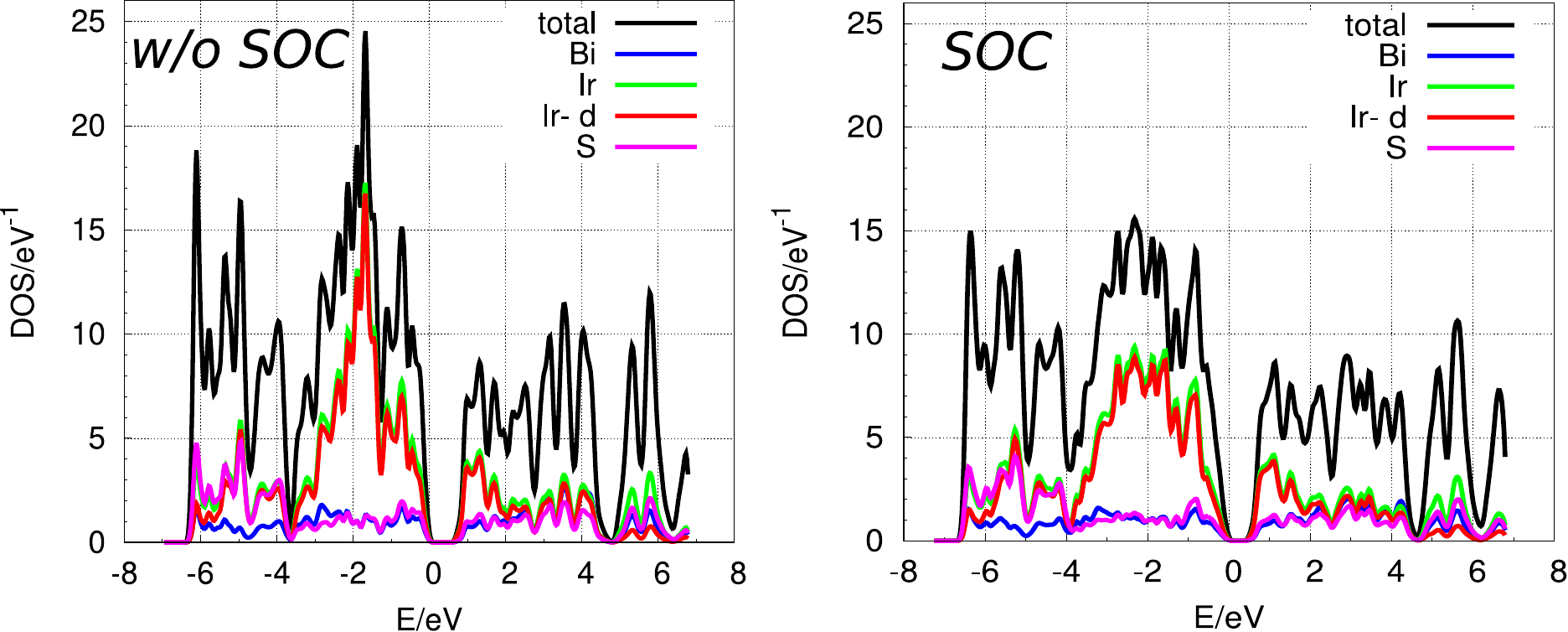}\caption{\label{fig:dos}The total and partial density of states in IrBiS without
(left) and with (right) SOC from the Wien2k calculations using GGA.
Explicitly we show the the Bi, Ir, S, and orbital resolved d states
for Ir.}
\end{figure}

As in the Se compound we observe a large $\sim0.34$ eV splitting
of the top-most valence band (in red) due to the SOC where the hole
pockets would appear upon light hole doping (indicated by a dotted
line with double headed arrow in Fig. \ref{fig:bands} (b)). These
bands have predominantly Ir-$d$ character (see Fig. \ref{fig:dos}).
The splitting vanishes at the time reversal symmetry points and the
Brillouin zone (BZ) boundaries and is almost negligible along the
BZ diagonal ($\Gamma$-R). The splitting also decreases as we move
away from the Fermi level.

\begin{figure}[h]
\includegraphics[width=0.7\columnwidth]{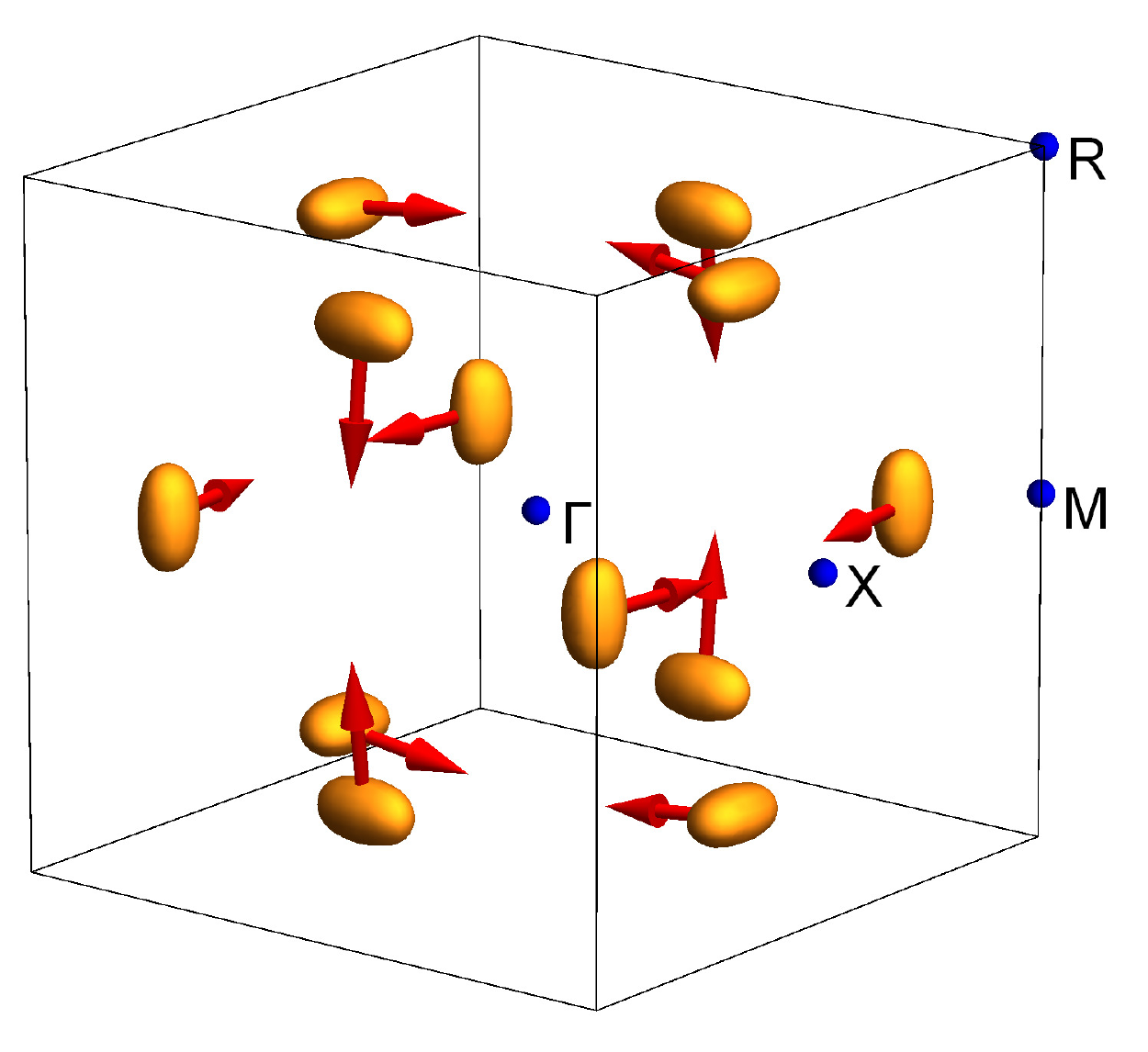} \caption{\label{fig:fermipockets} Hole pockets corresponding to the highest
valence band. The direction of spin in different pockets is indicated
by the red arrows.}
\end{figure}

The splitting \textit{at the top} of the valence band in IrBiS is
one order of magnitude larger than that observed \textit{at the top}
of the valence band\citep{cho2021observation,drichko2021dresselhaus}
in GaAs. \footnote{The maximal SOC splitting of the valence bands in GaAs and IrBiS are
of comparable magnitude, however, unlike to IrBiS, where the top of
the valence band is at the low symmetry point, the top of the valence
bands in GaAs is at the $\Gamma$ point where the splitting vanishes
due to the symmetry.} The valence band pockets in IrBiS are therefore, similarly to IrBiSe\citep{liu2020giant},
fully spin polarized as shown by the arrows in Fig. \ref{fig:fermipockets}.

While the intrapocket (intravalley) scattering should not be strongly
affected due to the spin polarization the interpocket (intervalley)
scattering is expected to be partially suppressed and anisotropic.
Each pocket has one corresponding pocket at the opposite momentum
with exactly the opposite spin polarization (due to the time reversal
symmetry), one pocket with nearly anti-parallel spin polarization,
one pocket with nearly parallel spin polarization and 8 pockets with
approximately perpendicular spin polarization. This could have profound
effects on spin relaxation and spin transport properties, however,
the detailed analysis of these effects is beyond the scope of the
present paper.

\subsection{Lattice vibrational modes and Raman spectra}

\begin{table}[p]
\caption{\label{tab:int_param}Calculated internal atomic parameters of IrBiS
after a complete structural relaxation using the PBE functional (without
SOC) for the relaxed lattice parameter, $a=6.213$ {Å}, and the
constrained lattice parameter, $a=6.143$ {Å} (experimental lattice
constant) compared to the experimental atomic positions as reported
in Ref. {[}\onlinecite{hulliger1963new}{]}. These parameters remain
similar also after the inclusion of SOC (not shown).\protect \\
}

\begin{ruledtabular}
\begin{tabular}{lcccccc}
atom & site (P2$_{1}$3) & x=y=z \footnote{$a=6.213$ {Å}} & x=y=z \footnote{$a=6.143$ {Å}} & x=y=z \footnote{expt, Ref. \citep{hulliger1963new}} &  & \tabularnewline
\hline 
Bi & 4a & 0.370 & 0.369 & 0.39 &  & \tabularnewline
Ir & 4a & 0.024 & 0.025 & 0.02 &  & \tabularnewline
S & 4a & 0.617 & 0.617 & 0.615 &  & \tabularnewline
\end{tabular}
\end{ruledtabular}

\end{table}

\label{subsec:phonons}

The calculated lattice parameters and the internal atomic parameters
obtained using the GGA functional are presented in Table~\ref{tab:int_param}
while Table~\ref{tab:phonon_modes} shows the phonon wavenumbers
calculated at the Brillouin zone '$\Gamma$' point. The phonon modes
symmetries according to the point group symmetry T(23) are also indicated.

The mode wavenumbers calculated using the relaxed lattice parameter,
$a=6.213$ {Å}, which is larger than the experimental, $a_{\mathrm{exp}}=6.143$
{Å}, appear significantly softer than the experimental ones. The
inclusion of SOC hardens most of the modes, but not significantly,
so the resulting wavenumbers remain $\sim5$\% to $\sim11$\% softer
than the experimental ones. The effect of SOC is the largest for the
three modes calculated to be around $\sim290$ cm$^{-1}$. The calculated\footnote{Without SOC.}
294.4 cm$^{-1}$ A mode softens by 1.9\% while the 275.9 cm$^{-1}$
E mode and the 282.6 cm$^{-1}$T mode harden by 3.5\% and 2.9\%, respectively.
All these modes are S ions displacements dominated (see Table SII
(Supporting Information)). Experimentally these modes lay between
305 cm$^{-1}$ (the E mode) and 311 cm$^{-1}$ (the T mode) with the
T mode being harder than A, contrary to the calculations.

Taking the smaller experimental lattice parameter hardens the modes
bringing most of the wavenumbers within a 3\% discrepancy band with
most of the modes remaining too soft. The inclusion of SOC slightly
improves match for most of the modes, however, the calculated\footnote{Without SOC.}
331.7 cm$^{-1}$ T mode wavenumber and the 306.9 cm$^{-1}$ A mode
wavenumber are significantly closer to the experimental wavenumbers
in the absence of SOC. In both cases they fall well within the 3\%
discrepancy band.

There are still a few modes that show discrepancies in excess of 3\%
even after inclusion of the SOC: the calculated\footnote{With SOC.}
84 cm$^{-1}$A mode, the 87.1 cm$^{-1}$ T mode, and the T mode pair
at 165.9 cm$^{-1}$ and 178.1 cm$^{-1}$. These modes show the largest
discrepancies also with the relaxed lattice parameter. Looking at
the order of modes, the highest wavenumber A mode is pushed below
the highest wavenumber E mode, contrary to what is observed in the
experiment.

It will be interesting to explore in future work whether going beyond
the pseudo-potential plane wave method and using the full potential
augmented plane wave method or using more advanced functionals to
treat the exchange correlation effects within the DFT would improve
the agreement.

For comparison we also computed the phonon modes at the $\Gamma$-point
in IrBiSe and found them to be systematically softer than in IrBiS.
The details are presented in Supporting Information.

\section{Summary and conclusions}

We measured phonon Raman spectra in IrBiS, a semiconductor which displays
a large bulk-type Dresselhaus valence bands splitting of $\sim0.3$
eV, and compared the experimental phonon wavenumbers to the \emph{ab
initio} predicted ones by taking into account the SOC effects.

Using two different excitation wavelengths all the symmetry predicted
phonon modes were experimentally observed. The calculated wavenumbers
appear systematically softer than the experimental ones. The inclusion
of SOC shows mostly a minor hardening of the phonon wavenumbers in
comparison to the lattice parameter effects, where constraining the
lattice parameter to the smaller experimental value hardens most of
the calculated phonon wavenumbers within a 3\% discrepancy band.

Comparing the phonon mode wavenumbers of the sister compound IrBiSe
to IrBiS we find a strong, $\sim35$ \%, softening of the chalcogen-displacements
dominated modes and a lesser, $\sim4$ \%, softening of the Ir/Bi-displacements
dominated modes.

The calculated electronic structure of IrBiS is found to be similar
to the sister compound IrBiSe with almost fully spin polarized valence-band
hole pockets.
\begin{acknowledgments}
The authors acknowledge the financial support of Slovenian Research
Agency (research core funding No-P1-0040 and P1-0044 and research
project funding J1-2458) for financial support.
\end{acknowledgments}

\bibliographystyle{JRS}

\begin{widetext}
\clearpage
\end{widetext}

\newpage
\phantom{a}
\newpage
\onecolumngrid
\begin{center}
{\large \bf Supplementary Information: \\
 Lattice vibrational modes in changchengite from Raman spectroscopy
and first principles electronic structure}\\
\vspace{0.3cm}
B. Chatterjee$^{1}$, D. Vengust$^{2}$, A. Mrzel$^{2}$, P.Sutar$^{2}$, E. Goreshnik$^{3}$, J. Mravlje$^{1}$, and T. Mertelj$^{2,4}$\\
$^1${\it Department of Theoretical Physics, Jozef Stefan Institute, Jamova
39, 1000 Ljubljana, Slovenia} \\
$^2${\it Department of Complex Matter, Jozef Stefan Institute, Jamova 39, 1000
Ljubljana, Slovenia} \\
$^3${\it Dept. of Inorganic Chemistry and Technology, Jozef Stefan Institute,
Jamova 39, 1000 Ljubljana, Slovenia}\\
$^4${\it Center of Excellence on Nanoscience and Nanotechnology Nanocenter
(CENN Nanocenter), Jamova 39, 1000 Ljubljana, Slovenia}\\
\end{center}

\section{Crystal characterization}

\begin{table}[p]
\caption{\label{tab:Composition}Composition of different crystals obtained
from Energy-dispersive X-ray spectroscopy.}

\begin{ruledtabular}
\begin{tabular}{cccc}
No. & Ir (at. \%) & Bi (at. \%) & S (at. \%)\tabularnewline
\hline 
1. & 32.8 & 33.5 & 33.8\tabularnewline
2. & 33.7 & 32.7 & 34.0\tabularnewline
3. & 31.0 & 33.6 & 35.4\tabularnewline
4. & 32.3 & 33.9 & 33.8\tabularnewline
\end{tabular}
\end{ruledtabular}

\end{table}

The results of from Energy-dispersive X-ray spectroscopy analysis
of different IrBiS crystals composition are shown in Table \ref{tab:Composition}.
Within the expected accuracy of the technique the composition Ir:Bi:S
is 1:1:1.

\section{SOC calculations treating the Bi 6p-1/2 state with local orbital}

\begin{figure*}[p]
\centering{}\includegraphics[width=1\textwidth]{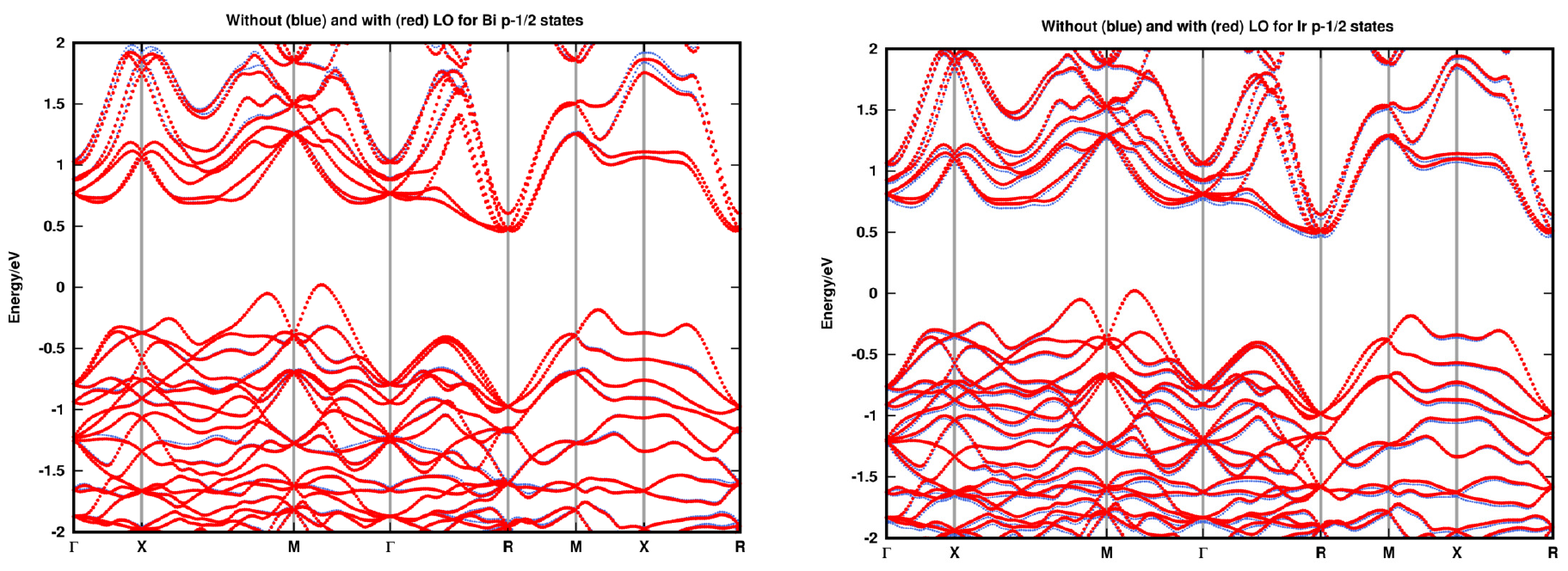}
\caption{\label{fig:bilo} Left: Bi $6p_{1/2}$ states treated with LO (red)
and without LO (blue). Right: Ir $6p_{1/2}$ states treated with LO
(red) and without LO (blue).}
\end{figure*}

To improve the description of the heavy elements $6p$ states one
can include the local orbitals (LO) as described in Ref. \cite{kunevs2001electronic}.
This was not included in the SOC results shown in the main paper,
however, to estimate the possible improvement of the calculations
we performed some calculations including the LO as well. A comparison
of the band structure when adding the heavier Bi atom LO shows no
qualitative and only negligible quantitative differences in the over
all band-structure as shown in Fig. \ref{fig:bilo}(left). The SOC
splitting of the top-most valence band is not at all affected after
inclusion of the LO. This is in agreement with the overall chemistry
of this compound where the states near the Fermi energy are dominated
by the Ir $d$ derived states. The other dominant characters are the
S $p$ states and Bi $p$ states with the S $p$ having higher contribution
on the occupied side, and the Bi $p$ on the unoccupied side. Hence,
the topmost valence band with the dominant Ir $d$ character is not
at all affected by adding Bi $6p_{1/2}$ LO.

We have further computed the band structure also adding the Ir $6p_{1/2}$
LO which also show just minor quantitative changes in the band structure
as shown in Fig. \ref{fig:bilo} (right).

\section{Lattice vibrational mode displacements in I\lowercase{r}B\lowercase{i}S}

In Table \ref{tab:phonon_displacements} we show calculated relative
phonon displacements magnitudes for each ionic species for the best
agreement case of constraining the lattice parameter to the experimental
value. Since for each species all sites are crystalographically equivalent
the displacement magnitudes must be equal across the species for a
single $q=0$ mode for the A and E symmetry modes\footnote{The complex E representations are from the point of view of the spatial
symmetry one dimensional.}. For the T symmetry modes the 4 symmetry equivalent unit cell sites
can have different displacement magnitudes so the range is given where
appropriate.

The high-frequency modes group is clearly S-ion displacements dominated
while the low frequency group shows rather mixed displacements with
somewhat larger contribution of the heaviest Bi ions in the four lowest
wavenumber modes.

The effect of the SOC inclusion on the relative wavenumber shift is
the largest for three S-ion-displacements dominated modes and the
145 cm$^{-1}$-mode that shows the largest relative Ir ion displacements.
The reason for this is not understood as the displacements of the
modes are rather complex and nonintuitive, due to the low-symmetry
distribution of the atoms within the unit cell.

The modes that show frequency discrepancies are all mixed modes and
do not show any outstanding relative displacement magnitudes pattern.

\begin{table*}[p]
\caption{\label{tab:phonon_displacements}Calculated ionic phonon displacements
magnitudes for the best agreement case of using the experimental lattice
parameter. A range is given (see text) for the T symmetry modes where
appropriate. The modes that show frequency discrepancies in excess
of 3\% are shown in magenta.}

\begin{ruledtabular}
\begin{tabular}{cccc>{\centering}p{0.8cm}cccccc}
\multicolumn{2}{c}{Experimental} & \multicolumn{2}{c}{Calculations} & Symmetry & \multicolumn{6}{c}{Displacement magnitude\footnote{Relative to the species with the maximum displacement for the given
mode.}}\tabularnewline
\midrule 
632.8 nm & 488 nm & $a=a_{\mathrm{exp}}$\footnote{with SOC} & $\Delta_{\mathrm{SOC}}$\footnote{Relative shift due to the SOC.} &  & \multicolumn{3}{c}{without SOC} & \multicolumn{3}{c}{with SOC}\tabularnewline
\midrule 
\multicolumn{3}{c}{Wavenumber (cm$^{-1}$)} & \% & - & \multicolumn{1}{c}{Ir} & \multicolumn{1}{c}{Bi} & \multicolumn{1}{c}{S} & \multicolumn{1}{c}{Ir} & \multicolumn{1}{c}{Bi} & \multicolumn{1}{c}{S}\tabularnewline
\midrule 
\textcolor{magenta}{84.4} & \textcolor{magenta}{-} & \textcolor{magenta}{81.3} & \textcolor{magenta}{-0.1} & \textcolor{magenta}{A} & \multicolumn{1}{c}{\textcolor{magenta}{0.49}} & \multicolumn{1}{c}{\textcolor{magenta}{1.00}} & \multicolumn{1}{c}{\textcolor{magenta}{0.51}} & \multicolumn{1}{c}{\textcolor{magenta}{0.47}} & \multicolumn{1}{c}{\textcolor{magenta}{1.00}} & \multicolumn{1}{c}{\textcolor{magenta}{0.49}}\tabularnewline
\textcolor{magenta}{93.0} & \textcolor{magenta}{93.7} & \textcolor{magenta}{87.1} & \textcolor{magenta}{1.0} & \textcolor{magenta}{T} & \textcolor{magenta}{0.74 - 0.83} & \textcolor{magenta}{0.81 - 1.00} & \textcolor{magenta}{0.59 - 0.66} & \textcolor{magenta}{0.74 - 0.81} & \textcolor{magenta}{0.83 - 1.00} & \textcolor{magenta}{0.59 - 0.65}\tabularnewline
97.9 & 98.2 & 96.4 & 1.4 & E & \multicolumn{1}{c}{0.94} & \multicolumn{1}{c}{1.00} & \multicolumn{1}{c}{0.34} & \multicolumn{1}{c}{0.92} & \multicolumn{1}{c}{1.00} & \multicolumn{1}{c}{0.34}\tabularnewline
121.6 & 122.0 & 118.3 & -0.1 & T & 0.48 - 0.51 & 0.82 - 1.00 & 0.12 - 0.18 & 0.48 - 0.51 & 0.81 - 1.00 & 0.10 - 0.15\tabularnewline
138.6 & 138.4 & 134.9 & 0.4 & A & \multicolumn{1}{c}{1.00} & \multicolumn{1}{c}{0.45} & \multicolumn{1}{c}{0.04} & \multicolumn{1}{c}{1.00} & \multicolumn{1}{c}{0.44} & \multicolumn{1}{c}{0.08}\tabularnewline
-- & 148.5 & 145.0 & 2.0 & T & 0.79 - 1.00 & 0.22 - 0.38 & 0.11 - 0.19 & 0.75 - 1.00 & 0.21 - 0.41 & 0.06 - 0.18\tabularnewline
169.0 & 168.7 & 164.4 & 0.9 & E & \multicolumn{1}{c}{1.00} & \multicolumn{1}{c}{0.90} & \multicolumn{1}{c}{0.66} & \multicolumn{1}{c}{1.00} & \multicolumn{1}{c}{0.88} & \multicolumn{1}{c}{0.70}\tabularnewline
\textcolor{magenta}{177.1} & \textcolor{magenta}{176.9} & \textcolor{magenta}{165.9} & \textcolor{magenta}{0.9} & \textcolor{magenta}{T} & \textcolor{magenta}{0.34 - 1.00} & \textcolor{magenta}{0.38 - 0.97} & \textcolor{magenta}{0.27 - 0.60} & \textcolor{magenta}{0.44 - 1.00} & \textcolor{magenta}{0.46 - 0.98} & \textcolor{magenta}{0.33 - 0.63}\tabularnewline
\textcolor{magenta}{188.1} & \textcolor{magenta}{188.0} & \textcolor{magenta}{178.1} & \textcolor{magenta}{1.0} & \textcolor{magenta}{T} & \textcolor{magenta}{0.64 - 1.00} & \textcolor{magenta}{0.70 - 0.94} & \textcolor{magenta}{0.17 - 0.21} & \textcolor{magenta}{0.67 - 1.00} & \textcolor{magenta}{0.71 - 0.94} & \textcolor{magenta}{0.18 - 0.21}\tabularnewline
304.8 & 305.4 & 310.0 & 3.3 & E & \multicolumn{1}{c}{0.09} & \multicolumn{1}{c}{0.03} & \multicolumn{1}{c}{1.00} & \multicolumn{1}{c}{0.09} & \multicolumn{1}{c}{0.03} & \multicolumn{1}{c}{1.00}\tabularnewline
307.5 & 307.0 & 301.8 & -1.7 & A & \multicolumn{1}{c}{0.04} & \multicolumn{1}{c}{0.06} & \multicolumn{1}{c}{1.00} & \multicolumn{1}{c}{0.04} & \multicolumn{1}{c}{0.06} & \multicolumn{1}{c}{1.00}\tabularnewline
311.2 & 310.7 & 313.8 & 2.4 & T & 0.08 - 0.11 & 0.02 & 0.40 - 1.00 & 0.08 - 0.09 & 0.02 - 0.03 & 0.84 - 1.00\tabularnewline
320.9 & - & 319.6 & -0.5 & T & 0.05 & 0.02 - 0.05 & 0.68 - 1.00 & 0.08 & 0.03 - 0.07 & 0.66 - 1.00\tabularnewline
328.8 & 328.3 & 335.1 & 1.0 & T & 0.05 & 0.06 - 0.08 & 0.94 - 1.00 & 0.05 & 0.04 - 0.05 & 0.74 - 1.00\tabularnewline
\end{tabular}
\end{ruledtabular}

\end{table*}

\section{$\Gamma$-point Phonon modes in I\lowercase{r}B\lowercase{i}S\lowercase{e}}

\begin{table}[p]
\caption{\label{tab:ph_mode_Se}Calculated phonon wavenumbers at the $\Gamma$-point
in the sister compound IrBiSe without SOC for the fully relaxed structure
with the theoretical lattice constant, $a=6.38$ {Å}.}
.
\begin{ruledtabular}
\begin{tabular}{ccccc}
IrBiS & IrBiSe &  &  & \tabularnewline
\multicolumn{2}{c}{Wavenumber (cm $^{-1}$)} & Symmetry & Activity & \tabularnewline
\hline 
-0.20 & 0.0 & T & I+R & \tabularnewline
79.04 & 75.43 & A & R & \tabularnewline
83.8 & 78.24 & T & I+R & \tabularnewline
91.2 & 87.8 & E & R & \tabularnewline
114.1 & 110.5 & T & I+R & \tabularnewline
127.6 & 124.5 & A & R & \tabularnewline
133.1 & 130.45 & T & I+R & \tabularnewline
154.1 & 140.81 & E & R & \tabularnewline
156.2 & 145.8 & T & I+R & \tabularnewline
165.9 & 158.5 & T & I+R & \tabularnewline
\hline 
275.9 & 189.8 & E & R & \multirow{5}{*}{S/Se dominated}\tabularnewline
294.4 & 184.5 & A & R & \tabularnewline
282.6 & 191.3 & T & I+R & \tabularnewline
300.8 & 199.4 & T & I+R & \tabularnewline
314.2 & 204.4 & T & I+R & \tabularnewline
\end{tabular}
\end{ruledtabular}

\end{table}

For comparison we also calculated the phonon mode wavenumbers at the
$\Gamma$-point (in the absence of SOC) in the sister compound IrBiSe.
We used similar settings and method like in the case of IrBiS. In
Tab. \ref{tab:ph_mode_Se} we present the phonon mode wavenumbers
in IrBiSe for a completely relaxed structure with the relaxed lattice
constant of $a=6.38$ {Å}, which is, similar as in IrBiS, found
larger than the experimental,\cite{hulliger1963new} $a_{\mathrm{exp}}=6.290$
{Å}.

There is an overall softening of the mode wavenumbers as compared
to IrBiS. While the low wavenumber modes (below $\sim160$ cm$^{-1}$)
are softer only by $\sim5$ \%, the high wavenumber Se modes (above
$\sim190$ cm$^{-1}$) are significantly softer by $\sim35$ \%. The
wavenumber of the doubly degenerate charge four Weyl phonon mode at
87.8 cm$^{-1}$ (2.633 THz) is consistent with the results presented
in Ref. {[}\onlinecite{liu2021charge}{]}. The corresponding doubly
degenerate mode in IrBiS is found at 91.2 cm$^{-1}$. It could be
an interesting future work to study the complete phonon dispersion
in IrBiS and further investigate the Weyl phonons.

\bibliographystyle{JRS}


\end{document}